\documentclass[a4paper,%
   a4paper,%
   BCOR12mm,%
   11pt,%
   abstracton,%
   pointednumbers,%
   tablecaptionabove,%
   footinclude,%
   halfparskip,%
   ]{scrartcl}
\usepackage[ilines]{scrpage2} 
\usepackage{epsfig}
\usepackage{multicol}
\usepackage{amsmath}
\usepackage[round,colon,sort,authoryear]{natbib}
\usepackage{multirow}
\usepackage{amsfonts}
\newcommand{\A}{\mathcal{P}_A} %Product A
\newcommand{\B}{\mathcal{P}_B} %Product B

\newtheorem{hypothesis}{Hypothesis}

\lehead{}
\rohead{\small\thepage}
\lofoot{\small kjhmagno\&jppabico/uplb/towards input device satisfaction...}
\cofoot{}
\rofoot{}

\begin{document}

\title{Towards Input Device Satisfaction Through\\Hand Anthropometry}

\author{Katrina Joy H. Magno and Jaderick P. Pabico\\
    \small{Institute of Computer Science}\\
    \small{University of the Philippines}\\
    \small{Los Ba\~{n}os, Laguna}\\
    \small{\tt\{kjhmagno, jppabico\}@uplb.edu.ph}
}
\date{}

\maketitle
\begin{abstract}
We collected the hand anthropometric data of 91 respondents to come up with a Filipino-based measurement to determine the suitability of an input device for a digital equipment, the standard PC keyboard. For correlation purposes, we also collected other relevant information like age, height, province of origin, and gender, among others. We computed the percentiles for each finger to classify various finger dimensions and identify length-specific anthropometric cut-points. We compared the percentiles of each finger dimension against the actual length of the longest key combinations when correct finger placement is used for typing, to determine whether the standard PC keyboard is fit for use by our sampled population. Our analysis shows that the members of the population with hand dimensions at extended position below 75th percentile and at 99th percentile are the ones who would most likely not reach the longest key combination for the left and the right hands, respectively. Using machine vision and image processing techniques, we automated the anthropometric process and compared the accuracy of its measurements to that of manual process'. We compared the measurement generated by our automated anthropometric process with the measurements using the manual one and we found out that they have a very minimal absolute difference. The data collected from this study could be used in other studies such as determining a good design for mobile and other handheld devices, or input devices other than keyboard. The automated method that we developed could be used to easily measure hand dimensions given a digital image of the hand and could be extended for measuring the entire human body for various other applications.
\end{abstract}

%\terms{Anthropometry}
%\keywords{hand anthropometry, digital image processing}
%=======================================================================
\section{Introduction}
%=======================================================================
% Why is this problem interesting? -=-=-=-=-=-=-=-=-=-=-=-=-=-=-=-=-=-=-
Among the many parts of the human body, the hand is unique in being free from repetitive locomotor duties that are usually done by the legs, eyelids, jaw, and other body parts with repetitive and habitual functions. In fact, the human hand is devoted entirely to functions of manipulation, which it does effectively due to the special anatomical configuration of its bones and muscles. The structural configuration of the human hand allows the opposing action of the thumb surface to the corresponding surfaces of the other four fingertips. Controlled by a highly evolved nervous system, the human is one among the few species on Earth known to be capable of executing a firm grasp~\citep{markze71,cutkosky90,begliomini08}. Because of this, the hand has become an indespensable part of the human body for controlling and effecting changes on the environment, either with or without the aid of tools.

The proper dimension of hand-controlled devices, such as computer input devices, is an important aspect of a product's design process. The user's impression about a product~$\A$ while she\footnote{We used the female gender as our writing style only. Depending on the context, we mean both or either gender, and not as a means to prejudice the opposite one.} is using it, in effect, determines her satisfaction level on~$\A$. In fact, studies~\citep{melcher03,wong05} show that the success of any product~$\B$ in the market depends solely on the satisfaction of the end user while using~$\B$, rather than on~$\B$'s technological development. Here, user satisfaction can be seen as the sum of a user's feelings and attitudes on factors affecting the use of~$\B$~\citep{bailey83}. Thus, the success or failue of any product is dependent on the end user's satisfaction. A key element affecting user satisfaction with a computer input device, such as a keyboard, is its ergonomics. The industrial and mechanical designers face a challenging task in  fitting the keys and other components into an appealing, ergonomic and durable product that is of the right size of the majority of the population of the target market.

A good design may increase the productivity of users, as well as their comfort, in using input devices. To be able to achieve satisfaction in using these devices, there is a need to come up with an ideal measurement that would not strain the human's body parts used to control the devices. Due to the race-based variability (induced by gene $\times$ environment interaction) in human's anatomical features, there is no one-size-fits-all device that one can create for the global population. Thus, designers must customize the device according to the dimensions of the body parts of the majority in a local population. A scientific way to do this is to conduct a meticulous and time-consuming process of surveying the lengths of the body parts of the human users. This scientific way of measuring the body lengths of human individual is called {\em anthropometry}, derived from the Greek words ``anthropo'' and ``metron'' that mean ``human'' and ``measure,'' respectively~\citep{ulijaszek94}. Based on our literature review, aside from few studies conducted for Filipinos~\citep{lu07,afinidad10,amongo10,novabos10,novabos12}, there is no anthropometric database that exists as of this writing that reflects the measurements of Filipino body parts for use in designing hand-held digital devices. In fact, most digital devices that are available in the Philippine market have been imported from other countries. In the absence a Filipino-based anthropometric database, these imports were designed with the anthropometric data, either of the population of the originating country at the worst, or that the importers used neighboring countries' data they assumed to have approximated the Filipino measurements at the best. In effect, most devices that enter the Philippines do not comfortably fit into the majority of the Filipinos' hands, sacrificing user comfort, usability, and safety.

\begin{figure}[t]
\centering\epsfig{file=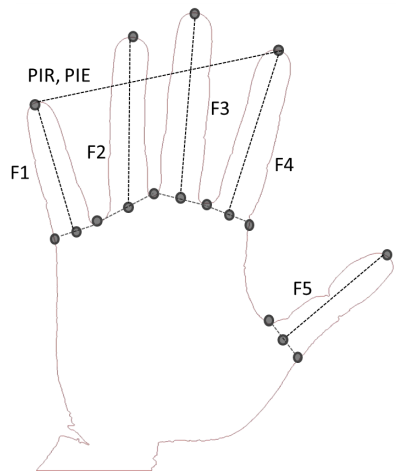, width=150px}
\caption{The seven linear measurements conducted for a hand, where the five linear measures are the five finger lengths and the other two are the linear measures between the fingertips of the pinky and the thumb when the hand is relaxed and extended. The codes F1, F2, F3, F4 and F5 correspond to the linear measures of the pinky, ring, middle, index and thumb fingers, respectively, and are defined in Table~\ref{tab:handDimension}. PIR and PIE are the codes for the linear measures between the tips of the pinky and the index when the hand is relaxed and  extended, respectively.}
\label{fig:measures}
\end{figure}

In this study, we conducted an anthropometric survey of the hands of 91 college students known to have come from different parts of the Philippines. Our purpose is to come up with an initial profile of users that may be used to design keyboards that will fit Filipino groups based on gender, age, and type of location of origin. To provide solution to the time, consistency and accuracy problems brought about by manual anthropometric process, we developed a computer-based anthropometry by combining techniques in machine vision and digital image processing in the hope of furnishing surveyors a fully-automated system that is fast, yet provides consistent measurements. Here, we obtained a total of 14 linear measurements, seven for each hand including the five fingers, and two hand positions used in conducting the key combination with the longest linear distance, as shown in Figure~\ref{fig:measures}. 

%-----------------------------------------------------------------------
%-----------------------------------------------------------------------

% what does this paper contain? -=-=-=-=-=-=-=-=-=-=-=-=-=-=-=-=-=-=-=-=
We present in this paper our contributions to the field of anthropometry:
\begin{enumerate}
\item A data supporting our claim that the manual hand anthropometric process, however strictly ``enforced,'' would still result in accuracy and consistency problems;
\item An automated hand anthropometric process, whose measurement is still within the surveyors' measurement, when the surveyors are not yet susceptible to human-induced errors; 
\item The anthropometric data of the hands of Filipinos based on gender, type of location of origin (i.e., rural or urban), age group, height, and weight; and
\item Mathematical equations that provide the various hand measurements as a function of the person's height.
\end{enumerate}

%=======================================================================
\section{Literature Review}
%=======================================================================
%what did others do? -=-=-=-=-=-=-=-=-=-=-=-=-=-=-=-=-=-=-=-=-=-=-=-=-=-

\subsection{Purpose of Hand Anthropometry}
%-----------------------------------------------------------------------
In practical design of devices, anthropometric surveys are usually done extensively in industries where devices must fit the workers hands for the purposes of increasing productivity, assuring comfort, and guarding the safety of the workers. Before a device is designed, anthropometric surveys are done on a group of people who share common characteristics, such as those belonging to specific locations, those with the same occupation, or those of the same gender or age group~\citep{hall06}. In most of these surveys, different body parts are measured depending on the survey's objectives~\citep{preedy12}. For example, the purpose of conducting an anthropometric survey of the hand is to aid in the optimization of hand-held devices, such as the one conducted for Jordanians by~\citet{nabeel08}. In this example, the authors presented the anthropometric results obtained from 120 female and 115 male adults from different cities in Jordan. They used some statistical tests to analyze the results and compared these with the results for other nationalities. Similar to one of our objectives, their data were aimed to influence the design of industrial tools imported to Jordan from other industrialized countries. 

\citet{dizmen12}, in a similar way, conducted a hand anthropometric survey for Hong Kong nationals where he used regression models to analyze his data. He recommended that the use of reference values for hand grip strength necessary in designing and fabricating industrial equipments would later prevent repetitive strain injuries to users. This translates, according to him, to a safer and healthier workplace. Additionally, hand measurements and their relation to the stature were studied by~\citet{zamila09} for the purposes of estimating the stature of Bengali nationals given the anthropometric measurements of the hand. 

Aside from industrial purposes, hand anthropometric data were also used to design hand-held devices like mobile phones with ``user satisfaction'' as one of the main objectives. In 2007, \citet{vimala07} determined that hand measurements and gender are factors in the user satisfaction of Short Message System (SMS) devices. Additionally, keypad sizes of mobile phones were collected and they found out that the females were more satisfied with the keypad sizes than the males. They recommended that the handheld devices must have larger keys and wider spaces between keys to accommodate those who have larger hand and finger sizes. The data they have collected could be used in designing mobile phone keypads and personalized mobile phone devices. 

Just like the work of~\citet{vimala07}, gender issue could also be one among the many goals of anthropometry. For example, \citet{amongo10} conducted an anthropometric survey where they measured a total of 38 body lengths of 284 women sampled in CALABARZON area. Their study was generally aimed to empower women in their participation to agricultural production processes. Their data is the first step toward ``gender equality'' because it establishes an anthropometric profile of CALABARZON women farmers, which in turn will serve as a database for future design and fabrication of gender-neutral agricultural machineries. 

%-----------------------------------------------------------------------
\subsection{Other Anthropometric Studies}
%-----------------------------------------------------------------------
There are many more applications of anthropometry that have been proven practical for some other fields of study. In the field of medicine, for example, quantitative comparison of anthropometric data with patients measurements before and after surgery provides the information needed by medical doctors for planning of and assessing the need of plastic and re-constructive surgery~\citep{farkas94}. In the domain of forensic anthropology, on the other hand, anthropometry-aided ``guessworks'' on likely measurements are important factors in the determination of individuals' appearance from their remains~\citep{rogers84,farkas94}. In the area of recovering missing children, anthropometric databases and the missing child's photographs are used in aging the child's appearance to infer what she would look like after so many years of being separated from her family~\citep{farkas94}. In the domain of entertainment computing, anthropometry is used to aid in the construction of face models for computer graphics applications, for creating life-like avatars in Computer-Generated Imagery (CGI)-based films, and for three-dimensional animation, among others~\citep{decarlo98}. 

There are many more examples of human-factors analyses and practical applications of anthropometry that can be found from the literature. For a comprehensive survey on these, the reader is directed to the work of~\citet{dooley82}. Most of the works surveyed here resulted into creating guides to help design products to fit most people. In general, however, anthropometric data is needed because it provides information on a range of enterprises that depend on knowledge of the distribution of measurements across human populations.

%-----------------------------------------------------------------------
\subsection{Problems in Manual Anthropometry}
%-----------------------------------------------------------------------
% what have we done? -=-=-=-=-=-=-=-=-=-=-=-=-=-=-=-=-=-=-=-=-=-=-=-=-=-
All of the above example studies used manual procedures in obtaining the hand measurements. The manual process takes quite a longer time because the anthropometric surveyors follow a strict and meticulous process (see~\citep{greiner92} and~\citep{NHANES07} for examples). Even though the process requires the human surveyors to be extremely careful and precise, the process is still realistically prone to human errors. If several surveyors are taking the measurement of a population sample, the surveyors themselves will introduce inconsistencies to their measurements. This is because one surveyor's standard and baselines for measurement can be different from another's, even if they meticulously follow the written process, and even under supervision. In this paper, we provide supporting data to this claim. We show that when five different surveyors $S_1, S_2, \dots, S_5$ measure the length of each of the ten fingers of one person $P_0$, at least 0.1 cm discrepancy is introduced in the database. Inconsistency arise when we do not know whose measurement is the correct one. But why hire multiple surveyors when hiring only one surveyor to conduct the survey for all the $n$~samples $P_1, P_2, \dots, P_n$ is an intuitive solution to the inconsistencies brought about by multiple surveyors? The answer to this that using one surveyor may also introduce accuracy problems. A surveyor who does the job of many, given a short amount of time, may get tired easily. Additionally, she may introduce short-cuts to the meticulous process so that she can measure more samples in a short amount of time. Again, we show in this paper the result of repeatedly measuring the finger lengths of one subject by one surveyor within two hours in 30-minute intervals. Here, inconsistencies were observed although some fingers were measured with the same value. The bottom line of all these is that using human surveyors, whether conducted solely by a lone surveyor or by many surveyors, has serious consistency and accuracy problems. Knowing that machines seldom get ``tired'' and can work at the same pace for long periods of time, automating the manual process should be the most logical solution to these problems.

\begin{figure}[t]
\centering\epsfig{file=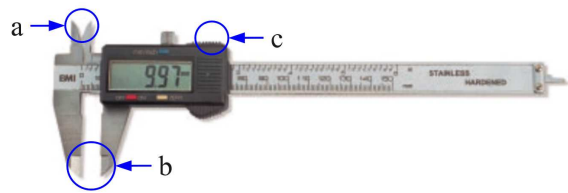, width=75mm}
\caption{An example digital caliper used for measuring (a) inside and (b) outside diameters, as well as the depth of a recess or hole. This portable measuring device looks like a slide rule with jaws and an LCD display. Many digital calipers now come with data output ports (c), so that the measurements can be read or logged by a computer.}
\label{fig:caliper}
\end{figure}

Semi-automating the anthropometric process is of course not a new solution. Instead of the surveyors eyeballing the measurements with the aid of a linear ruler (usually, a caliper or a ruled straight edge), some surveys used simple digital machines in order to automatically obtain the hand measurements. For example, \citet{nabeel08} used an electronic digital caliper (e.g., see Figure~\ref{fig:caliper}) to measure the twenty four hand dimensions included in their survey. However, the digital caliper is also manually-operated since it is just like a linear ruler but more accurate in measuring inside or outside diameters because it does not require a surveyor to perform an eyeball estimate. Therefore, a system that would provide full automation of the anthropometric process is needed so that the process of gathering anthropometric data is simplified, while the measurement process is both efficient and consistent.

%-----------------------------------------------------------------------
%-----------------------------------------------------------------------

%=======================================================================
\section{Problems in Manual Hand Anthropometry}
%=======================================================================
We claim in this paper that the manual hand anthropometric process, even though strict adherence to the meticulous procedure is followed by the surveyors under supervision, is still susceptible to accuracy and consistency problems. 

%-----------------------------------------------------------------------
\subsection{Accuracy Problem}
%-----------------------------------------------------------------------
The accuracy problem arises when $m > 0$~surveyors, $S_1, S_2, \dots, S_m$ perform the hand anthropometric survey on a sample population $\Phi$ of $n >> m$~subjects, where $\Phi=\{P_1, P_2, \dots, P_n\}$. Ordinarily, each surveyor~$S_i$ is assigned $n/m$ subjects and that the sub-population $\phi_i\subset\Phi$ of subjects assigned to $S_i$~is unique to her (i.e., $\Phi = \bigcup_i^m \phi_i$, $\emptyset = \bigcap_i^m \phi_i$ and $n = \sum_i^m |\phi_i|$). However, we contend that when all surveyors $S_i$, $\forall i = 1, \dots, m$, measure the same set $\pi\subset\Phi$ of $k$~subjects $\pi = \{P_1, P_2, \dots, P_k\}$, the measurements $m_{i,j}, \forall i=1,\dots, m$ will be different, $\forall j=1, \dots, m$. That is, the pairwise absolute difference between the measurements~$m_{x,j}$ and~$m_{y,j}$ of surveyors~$S_x$ and~$S_y$, respectively, on subject~$P_j\in\phi$ is statistically non-zero for some significance level $\alpha = 0.05$, $\forall x,y=1,2,\dots,m$, $x \not= y$. This means that $|m_{x,j} - m_{y,j}| > 0$, which implies that the surveyors are not accurate in measuring $P_j$.

%-----------------------------------------------------------------------
\subsection{Consistency Problem}
%-----------------------------------------------------------------------
Consistency problem arises when the surveyors measure the subjects in~$\pi$ repeatedly over a long period of time. That is, if a surveyor~$S_i$ repeatedly measures~$P_j\in\pi$ every $\Delta t$ within a time span of~$T$, her measurements $m_{i,j,t}$ will vary. This variability can be attributed to tiredness because the task was done during a long period of time~$T$, and boredom due to the repetitive nature of the task. Thus, if $|m_{i,j,t} - m_{i,j,t+\delta t}|>0$, $\forall t \in [0, T]$, then $S_i$ is not consistent in measuring~$P_j$ within~$T$. Note here that $S_i$ would have measured $P_j$ $\frac{T}{\Delta t} + 1$ times within~$T$.

%-----------------------------------------------------------------------
\subsection{A Two-factor Experiment}
%-----------------------------------------------------------------------
To test the claims that we made above, we let $m = 5$ surveyors manually measure the finger lengths of both hands of each of $n=91$ subjects following the meticulous process outlined in~\citet{greiner92}. For $T = 2$ hours, a set of $|\pi|=5$ subjects was repeatedly measured by the surveyors at an interval of $\Delta t = 30$ minutes, all of these while the surveyors are measuring the rest of the 86 other subjects once.  For each subject, the surveyors measure the lengths of the pinky, ring, middle, index, and thumb fingers of both hands, following the general measurement (interaction) model for each finger shown in Equation~\ref{eqn:model}. In the model, $\mu$ is the overall mean of the measurements averaged across $\Phi$, $S$, and $\tau$, while $P_j$ is the variability brought about by the $j$th subject, $S_i$ is the effect of the $i$th surveyor, $\tau$ is the effect of time on the measurement, and $\epsilon$ is the effect of a random error committed by the $i$th surveyor when measuring the $j$th subject at time~$t$.

\begin{equation}
   m_{i,j,t} = \mu + P_j + S_i + \tau_t + [S\times\tau]_{i,t} + \epsilon_{i,j,t}\label{eqn:model}
\end{equation}

In the above model, the term $[S\times\tau]_{i,t}$ depicts the interaction effect between the surveyors and the different times of measurement. We used a two-factor factorial analysis of variance (ANOVA) to test the following null hypothesis about the measurements of the subjects in~$\pi$:

\begin{hypothesis}
{\bf Accuracy-Consistency} - The pairwise difference between the respective mean measurements $m_{x,j,t}$ and $m_{y,j,t}$ of two surveyors $S_x$ and $S_y$ on subject $P_j\in\pi$ at time~$t$ is not different from zero at $\alpha = 0.05$. That is, $|m_{x,j,t} - m_{y,j,t}| = 0$, $\forall x,y = 1,2, \dots, m$,  $\forall j=1,2,\dots,|\pi|$ and $\forall t=1,2,\dots,T$. This means that the term $[S\times\tau]_{i,t} = 0$.
\end{hypothesis}

In case that the term $[S\times\tau]_{i,t} = 0$ at $\alpha = 0.05$, the above model reduces to an additive model, and the same ANOVA tests the following null hypotheses:

\begin{hypothesis}
{\bf Accuracy} - The pairwise difference between the respective mean measurements $m_{x,j}$ and $m_{y,j}$ of two surveyors $S_x$ and $S_y$ on subject $P_j\in\pi$ is not different from zero at $\alpha = 0.05$. That is, $|m_{x,j} - m_{y,j}| = 0$, $\forall x,y = 1,2, \dots, m$ and $\forall j=1,2,\dots,|\pi|$. Alternatively, it means $S_i = 0$.
\end{hypothesis}

\begin{hypothesis}
{\bf Consistency} - The pairwise difference between the mean measurements $m_{i,j,t_x}$ and $m_{i,j,t_y}$ of $S_i$ on subject $P_j\in\pi$ over time~$T$ is not different from zero at $\alpha = 0.05$. That is, $|m_{i,j,t_x} - m_{i,j,t_y}| = 0$, $\forall t_x \not= t_y$ and $\forall j=1,2,\dots,|\pi|$, which, in other words, means $\tau_t = 0$.
\end{hypothesis}

%-----------------------------------------------------------------------
\subsection{Result}
%-----------------------------------------------------------------------
\begin{table*}
\caption{Two-factor factorial ANOVA table of the right thumb lengths of five subjects as measured by five surveyors in a span of two hours. ``***'' means very highly significant at $\alpha = 0.001$, F means the computed F statistics, and Pr is the probability that the computed F  is greater than the critical F value.}\label{tab:anova}
\centering\begin{tabular}{l r r r r r}
\hline\hline
Source of Variation & Degree of & Sum of Squares & Mean Square & F & Pr > F\\
                    & Freedom\\
\hline
                    & \\
Subject ($P_j$)     &     4     & 1.07  & 0.27 & 105.74 & < 0.0001$^{***}$\\
Surveyor ($S_i$)    &     4     & 0.37  & 0.09 &  36.66 & < 0.0001$^{***}$\\
Time ($\tau_t$)     &     4     & 0.45  & 0.11 &  44.58 & < 0.0001$^{***}$\\
$[S\times\tau]_{i,t}$ &  16     & 0.56  & 0.03 &  13.75 & < 0.0001$^{***}$\\
Error ($\epsilon_{i,j,t}$) &  96     & 0.24  & 0.002 \\
\hline
Total                 & 124     & 2.69 \\
\hline\hline
\end{tabular}
\end{table*}
%-----------------------------------------------------------------------

Table~\ref{tab:anova} shows the ANOVA table for the right thumb length as measured by surveyors $S_1, S_2, \dots, S_5$ over a course of $T=2$ hours. Without loss of generality, it suffices to show that both accuracy and consistency problems exist in the manual hand anthropometry if only one finger exhibit the rejection of the null hypothesis that $[S\times\tau]_{i,t} = 0$. Although the ANOVA of all fingers exhibit a similar result as that shown in Table~\ref{tab:anova}, we will only show and discuss the result for the right thumb. We chose the thumb because it is one of the fingers involved in the difficult key combinations when using the standard keyboard, while the right hand was chosen because most of those that we surveyed are right-handed. Based on the table, the interactive effect factor  $[S\times\tau]_{i,t}$ is highly significantly different from zero at $\alpha = 0.001$. This means that the surveyors themselves have mean measurements that are different from each other for the right thumb at any time~$t$. Having the factor  $[S\times\tau]_{i,t} \not= 0$ means that the surveyors have introduced the accuracy and consistency problems in their measurements.

%-----------------------------------------------------------------------
\begin{figure*}
\centering\epsfig{file=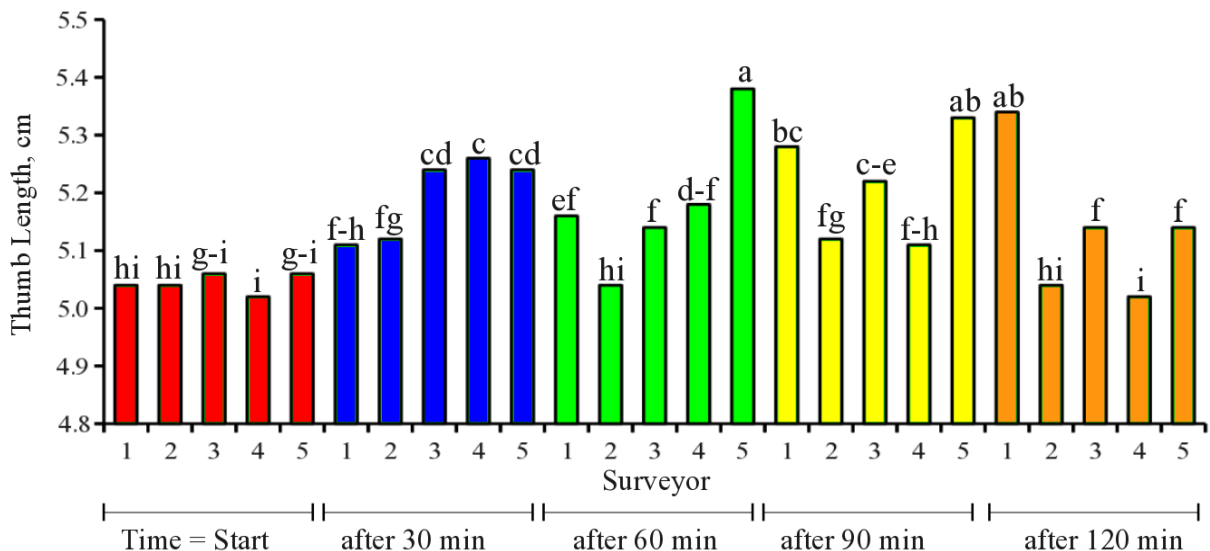, width=450px}
\caption{The comparison of the mean measurements in centimeters of the right thumb finger as measured by five surveyors $S_1, S_2, \dots, S_5$ at different time $t = \{0, 30, 60, 90, 120\}$ (minutes). Bars of the same color are mean measurements at the same time~$t$ while bars with the same letters have mean measurements that are not significantly different from each other at $\alpha=0.05$ by DMRT.}
\label{fig:dmrt}
\end{figure*}
%-----------------------------------------------------------------------

Figure~\ref{fig:dmrt} shows a visualization of the ANOVA. To better visualize the variance brought about by different surveyors, as well as by different time, we conducted a pairwise comparison of mean measurements using the multiple range test (DMRT) developed by~\citet{duncan55}. Each bar in the figure corresponds to the mean measurement in centimeters by surveyor~$S_i$, $\forall i = 1, 2, \dots, 5$. The bars with the same color are the measurements conducted at the same time such that all red bars were measured at time $t=0$, all blue bars at time $t=30$ minutes, and the subsequent similarly colored bars at a time interval of 30 minutes thereafter. The letters on each bar corresponds to the group of mean measurements where the differences of the measurements are not different from zero at~$\alpha=0.05$ according to DMRT. Differently grouped mean measurements are those whose differences are significantly different from zero at the same~$\alpha$. 

As seen in the figure, all red bars belong to the ``i'' group which means that the respective measurements of all surveyors did not vary significantly with each other at time $t=0$. After $t=30$ minutes, except for $S_1$, all surveyors have significantly different measurement from theirs at $t=0$ owing to the higher groups that their respective measurements belong according to DMRT. Here, $S_1$ and $S_2$ have statistically the same measurements, which are different from that of $S_3$'s, $S_4$'s, and $S_5$'s. At $t=60$ minutes, except for $S_2$, all surveyors have significantly different measurements from their measurements at $t=0$, while $S_1$, $S_3$, and $S_4$ have measurements that are statistically the same with each other and at the same time different from that of $S_1$'s and $S_5$'s. At $t=90$ minutes, $S_1$ and $S_3$ exhibited the same measurements, as well as that of $S_1$ and $S_5$, but not between $S_3$ and $S_5$. $S_2$ and $S_4$, likewise, exhibited the same measurements. At $t=120$ minutes, the measurements of $S_2$ and $S_4$ are not statistically different from their respective measurements at $t=0$. $S_3$ and $S_5$ have exactly the same measurements, but are statistically different from that of $S_2$'s and $S_4$'s. The measurement of $S_1$ is the highest among the five at this time, and is significantly different from all of them.

Doing the same DMRT analysis for all the fingers, we found out that all mean measurements at $t=0$ are grouped together at $\alpha=0.05$, although we are not showing the visualization here for the interest of space. We found out, however, that the surveyors at $t=0$ exhibit accurate measurements. However, at all $t>0$, the measurements of the surveyor on each finger have significant variability. We speculate here that the variability is brought about by the human factors in performing manual tasks, such as boredom due to the repetitive nature of the task, as well as tiredness, because the task has been done by the surveyors for a long time. At $t=0$, we can safely say that the surveyors are not yet affected by boredom and tiredness, and so their respective measurements are still statistically accurate. Thus, we can use these measurements as a basis to compare the result of our automated system.

%=======================================================================
\section{Hand Anthropometry}
%=======================================================================

\subsection{Demography of Respondents}
%-----------------------------------------------------------------------
We considered a total of $n=91$ tertiary students from a national university, where most students generally came from different regions in the Philippines. Because of this, they were also asked about their places of origin or the province where they came from, and whether their places of origin is considered of urban or of rural type, aside from the standard information such as age, gender, weight and height. Asking the respondents about the type of their places of origin is important in determining the general characteristics of the measurements of their hands because of the obvious differences in the rural and urban lifestyles. Students in the rural area are generally considered to have larger or longer hands due to exposure to manual work compared to those in urban area who are considered to have limited exposure to manual tasks.

We randomly selected these students from various groups whom we know to use computers most of the time. Although most respondents were BS Computer Science students, because they used computers in almost all of their subjects, we have also selected students from other courses that had computer laboratories in one or more of their subjects. The sampled population composed of mostly sophomore, junior, and senior students. Upon obtaining informed consents from all the respondents, we measured the finger dimensions of each hand of 41 male and 50 female students following a meticulous and strict procedure similar to the one outlined in~\citet{greiner92}. 

Table~\ref{tab:urbanRural} summarizes the number of respondents by gender and by type of places of origin. There were 41 male and 50 female respondents, while there were 52 who came from an urbanized place and 39 from a rural place. Of the 52 urban dwellers, 20 are male and 32 are female. Among the 39 rural folks, 21 are male and 18 are female. Table~\ref{tab:informationSummary} shows the mean, median, standard deviation and range of age (in years), height (in cm), and weight (in Kg) of the respondents. The respondents are generally in their late teens with an average of about 18 years and ranges from 17 to 21 years old. Their stature ranges from 148 to 163 cm with an average height of about 162 cm. They have a an average weight of 54.22 Kg with a spread that ranges from 30 Kg to 90 Kg.
%-----------------------------------------------------------------------
\begin{table}[t]
\centering
\caption{Number of respondents by gender, by type of place of origin, and by age group.}
\label{tab:urbanRural}
\begin{tabular}{| c | c c | c |}
\hline\hline
Category& \multicolumn{2}{c|}{Gender} & Total\\
\cline{2-3}
        & Male & Female & \\
\hline
Urban & 20 & 32 & 52\\
Rural & 21 & 18 & 39\\
\hline
Total & 41 & 50 & 91\\
\hline
\hline
Early Teens & 29 & 44 & 73 \\
Late Teens  & 12 & 6 & 18\\
\hline
Total & 41 & 50 & 91\\
\hline\hline
\end{tabular}
\end{table}
%-----------------------------------------------------------------------
\begin{table*}[t]
\centering
\caption{Summary statistics of the age, height and weight of respondents.}
\label{tab:informationSummary}
\begin{tabular}{|l|c c c c|}
\hline\hline
Characteristics & Mean & Median & Standard & Range\\
                &      &        & Deviation & \\
\hline
Age (year) & 18.03 & 18.00 & 0.75 & 17-21\\
Height (cm) & 161.71 & 161.54 & 8.45 & 148-163\\
Weight (kg) & 54.22 & 51.00 & 11.94 & 30-90\\
\hline\hline
\end{tabular}
\end{table*}
%-----------------------------------------------------------------------
\subsection{Manual Measurement}
%-----------------------------------------------------------------------
We manually measured the respective lengths of the five fingers of each hand using a graduated straight rule. We measured the finger lengths by placing the ruler at the approximate middle of the finger along the finger length. We defined each finger length as summarized in Table~\ref{tab:handDimension}. 
%-----------------------------------------------------------------------
\begin{table*}[t]
\centering
\caption{Finger length for each hand and the corresponding code.}
\label{tab:handDimension}
\begin{tabular}{|l l p{10cm}|}
\hline\hline
Code & Finger & Definition\\
\hline
F1 & Thumb & \parbox{10cm}{From the tip of the thumb to the line found at the root of the finger when palm is facing up, measured for both left and right hands}\\
\hline
F2 & Index & \parbox{10cm}{From the tip of the index finger to the line found at the root of the finger when palm is facing up, measured for both left and right hands}\\
\hline
F3 & Middle & \parbox{10cm}{From the tip of the middle finger to the line found at the root of the finger when palm is facing up, measured for both left and right hands}\\
\hline
F4 & Ring & \parbox{10cm}{From the tip of the ring finger to the line found at the root of the finger when palm is facing up, measured for both left and right hands}\\
\hline
F5 & Pinky & \parbox{10cm}{From the tip of the pinky finger to the line found at the root of the finger when palm is facing up, measured for both left and right hands}\\
\hline\hline
PIR & Pinky--Index (relaxed) & \parbox{10cm}{From the tip of the pinky finger to the tip of the index when the hand is relaxed, measured for both left and right hands}\\
\hline
PIE & Pinky--Index (extended) & \parbox{10cm}{From the tip of the pinky finger to the tip of the index when the fingers are extended, measured for both left and right hands}\\
\hline\hline
\end{tabular}
\end{table*}
%-----------------------------------------------------------------------
We also measured, for both hands, the linear length created by a combination of two fingers used by human typists when pressing the longest two-key combinations. A two-key combination is done by pressing two keys of the keyboard at the same time. These key combinations are the SHIFT--5 key combination for the left hand, and the SHIFT--6 key combination for the right hand. Following the standard finger assignments for keys (Figure~\ref{fig:keys}), the fingers used to perform these combinations for either hand are the pinky and the thumb. For each hand, we measured two linear lengths of the pinky--thumb combination: (1) when the two fingers are relaxed, and (2) when the two are extended, which are the PIR and PIE in Table~\ref{tab:handDimension}. In PIR, the user is at her most comfortable when conducting the key-combination with the longest linear length, while PIE is when she is straining too much just to make the combination. This is of course assuming that the key combination can be made by at least the PIE. Intuitively, it is much more difficult for a person to conduct the key combination when even the PIE is not enough to make it.

%-----------------------------------------------------------------------
\subsection{Acquisition and Preprocessing of Hand Images}
\begin{figure}[htb]
\centering\epsfig{file=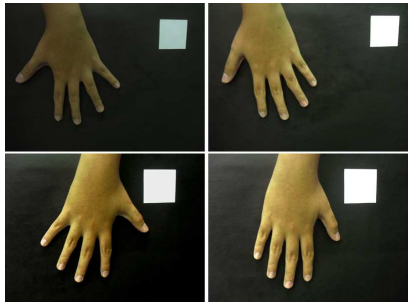, width=150px}
\caption{Hand images captured using a digital camera clockwise from top left (a)~extended left, (b)~relaxed left, (c)~extended right, and (d)~relaxed right. Notice the white square reference on each image.}
\label{fig:handsample}
\end{figure}

%-----------------------------------------------------------------------
After we manually measured the finger lengths, we captured the hand images using a digital color camera, with both relaxed and extended fingers. Thus, we collected a total of four hand images per respondent. Figure~\ref{fig:handsample} shows examples of the hand images of one respondent. In each color image $I_c$, we included a $5\times 5$ cm$^2$ square which we used as our reference for computing the actual finger dimensions. We also used this reference square for transforming the image (e.g., scaling and rotating) when the reference is not a perfect square due to differences in camera angles and other unaccounted variances during the image acquisition. Given the pixel coordinates $(x_{ul}, y_{ul})$, $(x_{ur}, y_{ur})$, $(x_{lr}, y_{lr})$ and $(x_{ll}, y_{ll})$ of the upper left, upper right, lower right and lower left corners, respectively, of the reference square, the reference is a perfect square if and only if $x_{ul} = x{ll}$, $x_{ur} = x_{lr}$, $y_{ul} = y_{ur}$, $y_{ll} = y_{lr}$, and $x_{ur} - x_{ul} = x_{lr} - x_{ll} = y_{ur} - y_{lr} = y_{ul} - y_{ll}$.

%-----------------------------------------------------------------------
%-----------------------------------------------------------------------

We converted the $I_c$ into gray-scale images $I_g$, and then we applied the median filter to $I_g$ to remove the noise. We then converted $I_g$ into a binary image $I_2$ using thresholding. We applied erosion to $I_2$ to further remove the noise brought about by $I_g$--$I_2$ conversion, while at the same time, we applied dilation twice so that the shape of the foreground object (i.e., the hand and the reference square) is, as much as possible, retained.

%-----------------------------------------------------------------------
\subsection{Automated Measurement from Images}
\begin{figure}[h]
\centering\epsfig{file=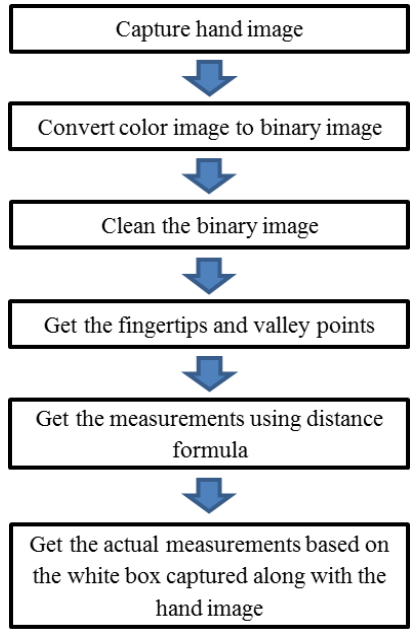, width=150px}
\caption{The flowchart for the automatic measurement of hand dimensions is a linear process composed of six steps.}
\label{fig:flowchart}
\end{figure}

%-----------------------------------------------------------------------
Our discussion of the automated process will be aided visually by Figure~\ref{fig:flowchart}, which shows our procedure for automatically obtaining the finger measurements from~$I_2$. To be able to get the lengths of each finger defined in Table~\ref{tab:handDimension}, we have to locate first the extrema points (i.e., fingertips and valley points) using~$I_2$. From~$I_2$, we determined the coordinates of the boundary of the foreground image using a simple row-wise horizontal scan. We saved these coordinates in an array $C = \{(x,y)\in\mathbb{Z}^{+}\}$, arranged in such a way that the profile of the foreground is as if being traced index-wise by~$C$.

\begin{figure}[t]
\centering\epsfig{file=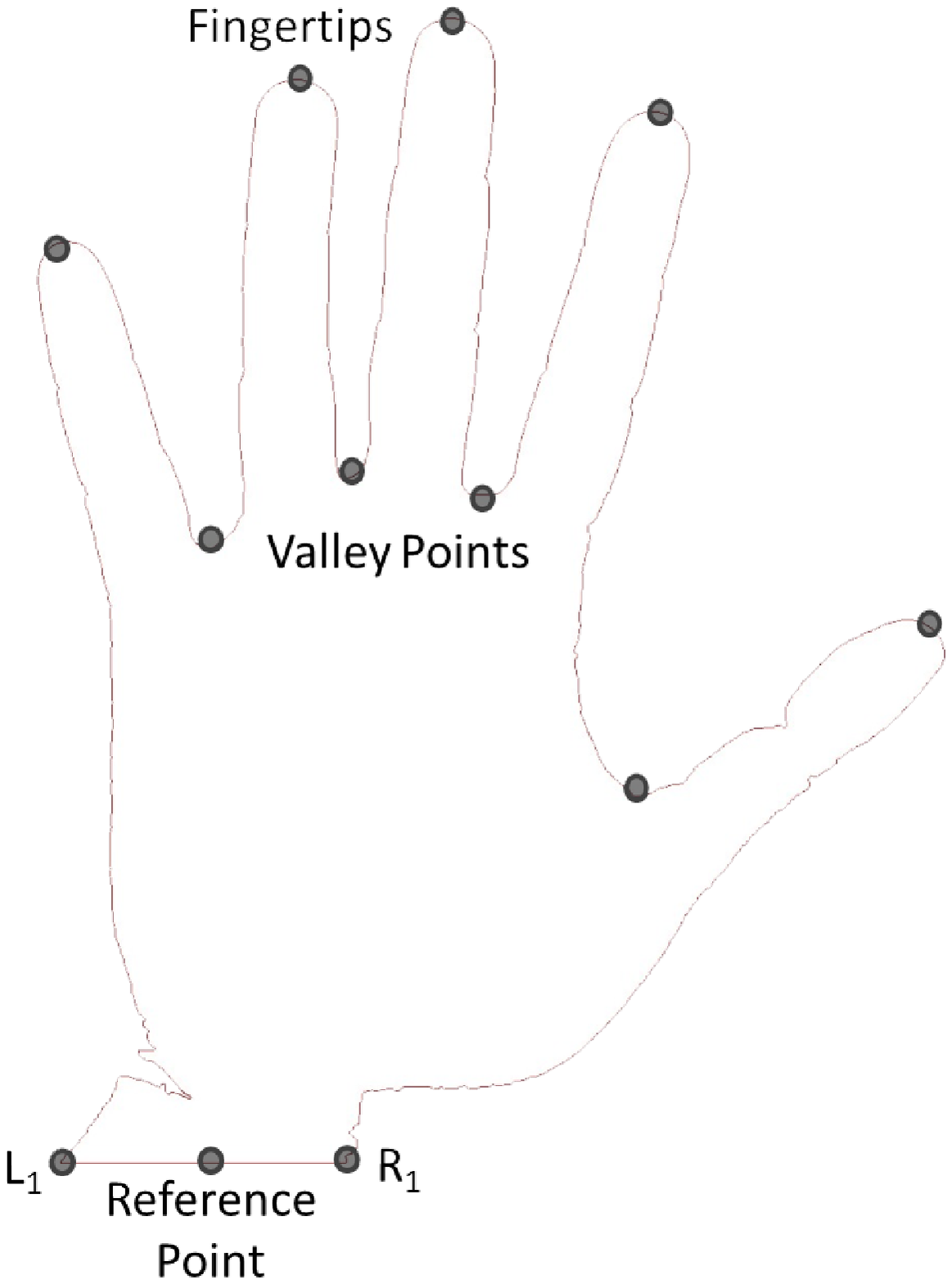, width=150px}
\caption{Fingertips and valley points of the hand image}
\label{fig:fingertips}
\end{figure}

We selected a reference point $C_r = (x_r, y_r)$ along the wrist as shown in Figure~\ref{fig:fingertips} by computing for it using the first and last array elements of~$C$ as follows $C_r = (\lceil\frac{x_1 + x_{|C|}}{2}\rceil, \lceil\frac{y_1 + y_{|C|}}{2}\rceil)$. For all points $C_i\in C$ (i.e., $\forall i=1,2,\dots,|C|$), and in the order of the index~$i$, we computed for the cartesian distance~$D_i $between $C_r$ and $C_i$: $D_i = \sqrt{(x_r - x_i)^2 + (y_r - y_i)^2}$. Our theoretical basis for marking the five fingertips and four valley points is to analytically compute for all $i$s where $\frac{d D}{d i} = 0$. However, this is not possible because we do not know the function that describes $D_i$. However, we did it by numerically choosing all indexes~$i$ when $|D_i - D_{i-1}| < \sigma$, where $0 < \sigma < 1$ is some threshold value. In this numerical method, we first found a fingertip (either that of the thumb or the pinky), followed by a valley point, and then another fingertip, in that order.

\begin{figure}[t]
\centering\epsfig{file=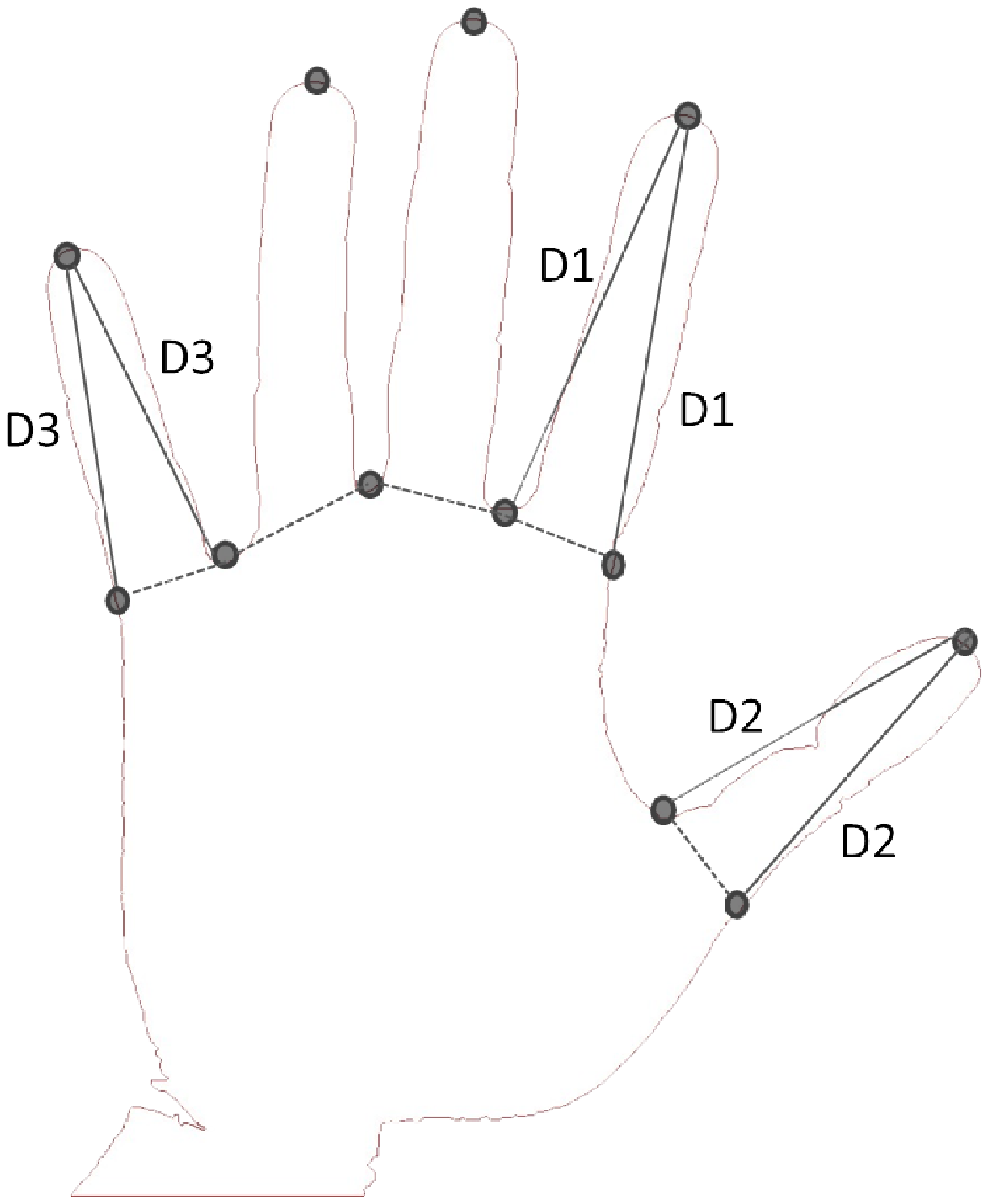, width=150px}
\caption{Fingertips and valley points of the hand image}
\label{fig:baselines}
\end{figure}

We needed the valley points to determine the baseline for each finger. However, from those four valley points, we can only determine the baselines for two fingers: the middle and the ring fingers. The  thumb, index, and pinky fingers do not have an opposing valley point, respectively. For the fingers that do not have an opposing valley point, we found a point along the profile of the hand that will form an isosceles triangle with the valley point and the finger tip (see Figure~\ref{fig:baselines}). We then determined the midpoint of these baselines after having determined the baselines of all fingers. The distance between and fingertip and baseline midpoint is the approximated finger length. Using a simple ratio and proportion, we computed for the actual length of the finger using the dimensions of the reference square.

The program was run and the hand dimensions of the finger in both left and right hands were recorded and summarized. A Table~\ref{tab:comparison} summarizes the mean and standard deviation of the actual hand measurements and the output of the program. The mean absolute difference between the actual hand measurements and the program output was computed by getting the average of the absolute differences of the actual data and the program output, as in equation~\ref{eq:mean}. The mean of the actual measurements and the measurements given by the program for the fingers differ by at most 0.73 centimeters. 

\begin{equation}
\label{eq:mean}
MD = \frac{\sum{|y2 - y1|}}{n}
\end{equation}

In the equation above, \begin{math}y_{1}\end{math} is the actual measure, \begin{math}y_{2}\end{math} is the measure given by the program and \begin{math}n\end{math} is the number of samples.

\begin{table*}[t]
\centering
\caption{Comparison between the actual hand measurements and the output of the automated system.}
\label{tab:comparison}
\begin{tabular}{| p{2cm} | p{2cm} p{2cm} | p{2cm} p{2cm} | p{2cm} |}
\hline
\parbox{2cm}{Hand\\Dimension} & \parbox{2cm}{Mean\\(actual)} & \parbox{2cm}{SD\\(actual)} & \parbox{2cm}{Mean\\(program)} & \parbox{2cm}{SD\\(program)} & \parbox{2cm}{Mean\\Absolute\\Difference} \\
\hline
\hline
\multicolumn{6}{|c|}{Right}\\
\cline{1-6} 
F1 & 5.95	& 0.57 & 5.48 & 0.61 & 0.55\\
F2 & 6.92 & 0.75 & 	7.22 &	0.66 &	0.53 \\
F3 & 7.73 &	0.47 &	8.05 &	0.69 &	0.47\\
F4 & 7.12 & 0.55 &	7.48 &	0.63 &	0.42 \\
F5 & 5.68 &	0.51 &	5.56 &	0.63 &	0.37 \\
\hline
\hline
\multicolumn{6}{|c|}{Left}\\ 
\cline{1-6} 
F1 & 5.95 &	0.63 &	5.28 &	0.67 &	0.73 \\
F2 & 7.02 &	0.48 &	7.43 &	0.95 &	0.62 \\
F3 & 7.74 &	0.49 &	7.40 &	0.83 &	0.51 \\
F4 & 7.05 &	0.85 &	8.06 &	0.72 &	0.49 \\
F5 & 5.71 &	0.48 &	5.36 &	0.78 &	0.53\\
\hline
\end{tabular}
\end{table*}

%=======================================================================
\section{Anthropometric Data of Filipinos}
%=======================================================================
We discuss in this section the percentile values (1st, 5th, 25th, 50th, 75th, 95th and 99th) of the finger lengths (F1, F2, $\dots$, F5), as well as the PIR and PIE of both hands of the surveyed population. As suggested by~\citet{nabeel08}, the extreme percentiles (1st, 5th, 95th and 99th) are particularly helpful statistics because they influence the ease of use and the suitability design parameters of the equipments. He warned, however, that the hand measurements that fall in these percentile ranks should carefully be used. For example, if the design of the input device is leaning towards those who are in the lower percentile, the device design might be too small for those who are in the middle and high percentiles. However, if the design is leaning towards the higher percentile ranks, those in the lower rank might not be able to use the input device at all since the size might be too big for them. PIR and PIE are helpful measures in determining the sizes of each key in the keyboard and the total dimension of the keyboard itself since this determines if the user will be able to reach the keys, especially if key-combinations are needed to be pressed when typing.

\begin{figure*}[t]
\centering\epsfig{file=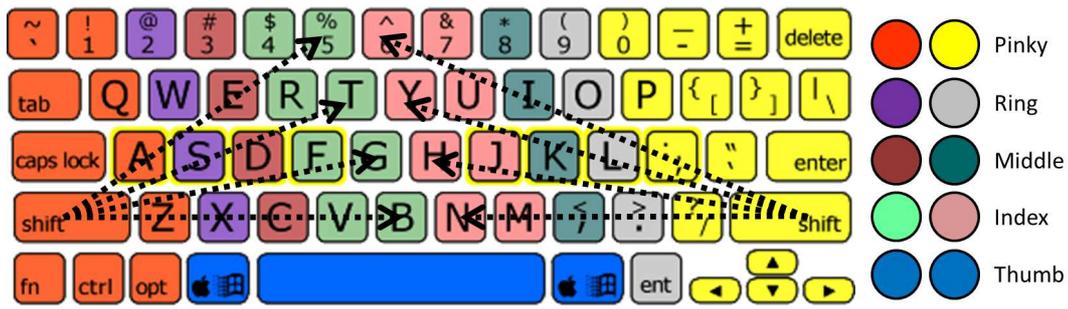, width=400px}
\caption{Correct keyboard hand placement and key combinations with the longest range}
\label{fig:keys}
\end{figure*}

Table~\ref{tab:keyboard} shows the actual dimension of the keyboard key combinations with the longest range. We obtained the dimensions by measuring using a ruled straight edge the linear distance between the respective centers of the keys in a standard 45cm-wide desktop keyboard. This is also illustrated in Figure~\ref{fig:keys}.

\begin{table*}[t]
\centering
\caption{Keyboard Key Combinations with longest range and their distances}\label{tab:keyboard}
\begin{tabular}{|l c c|}
\hline
Key Combinations & Fingers Used & Distance (cm) \\
\hline
\hline
Shift + 5 & Left Pinky + Index & 13\\
Shift + T & Left Pinky + Index  & 12.7\\
Shift + G & Left Pinky + Index & 12.5\\
Shift + B & Left Pinky + Index & 12.0\\
Shift + 6 & Right Pinky + Index & 18.0\\
Shift + Y & Right Pinky + Index & 16.5\\
Shift + H & Right Pinky + Index & 15.5\\
Shift + N & Right Pinky + Index & 14.5\\
\hline
\end{tabular}
\end{table*}

%-----------------------------------------------------------------------
\subsection{All Samples}

\begin{figure}[t]
\centering\epsfig{file=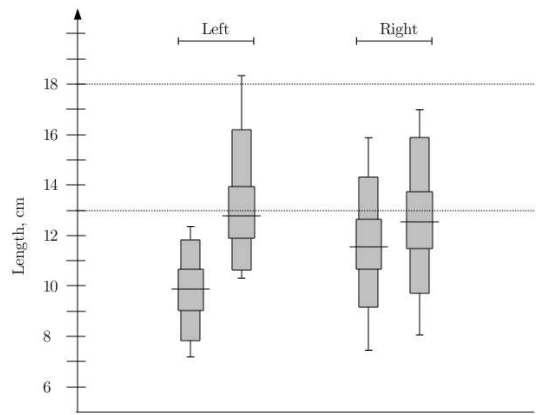, width=70mm}
\caption{Double-box and whisker plot of left and right PIR and PIE of the whole population.}
\label{fig:data-all}
\end{figure}

\begin{figure}[t]
\centering\epsfig{file=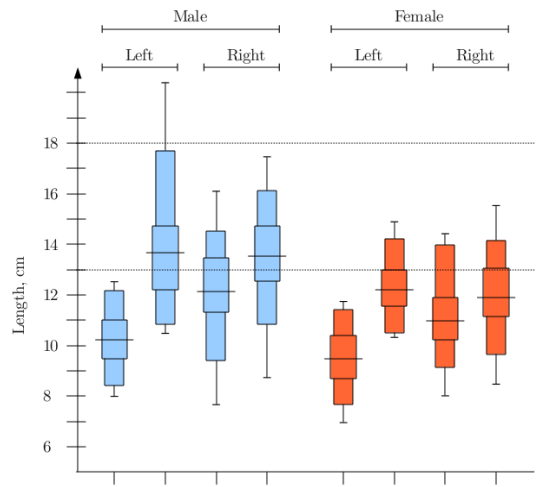, width=70mm}
\caption{Double-box and whisker plot of left and right PIR and PIE of the population by gender.}
\label{fig:data-gender}
\end{figure}

\begin{figure}[t]
\centering\epsfig{file=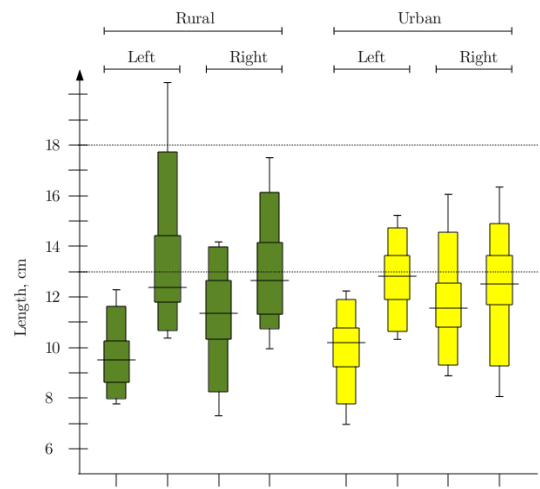, width=70mm}
\caption{Double-box and whisker plot of left and right PIR and PIE of the population by type of location of origin.}
\label{fig:data-location}
\end{figure}

\begin{figure}[t]
\centering\epsfig{file=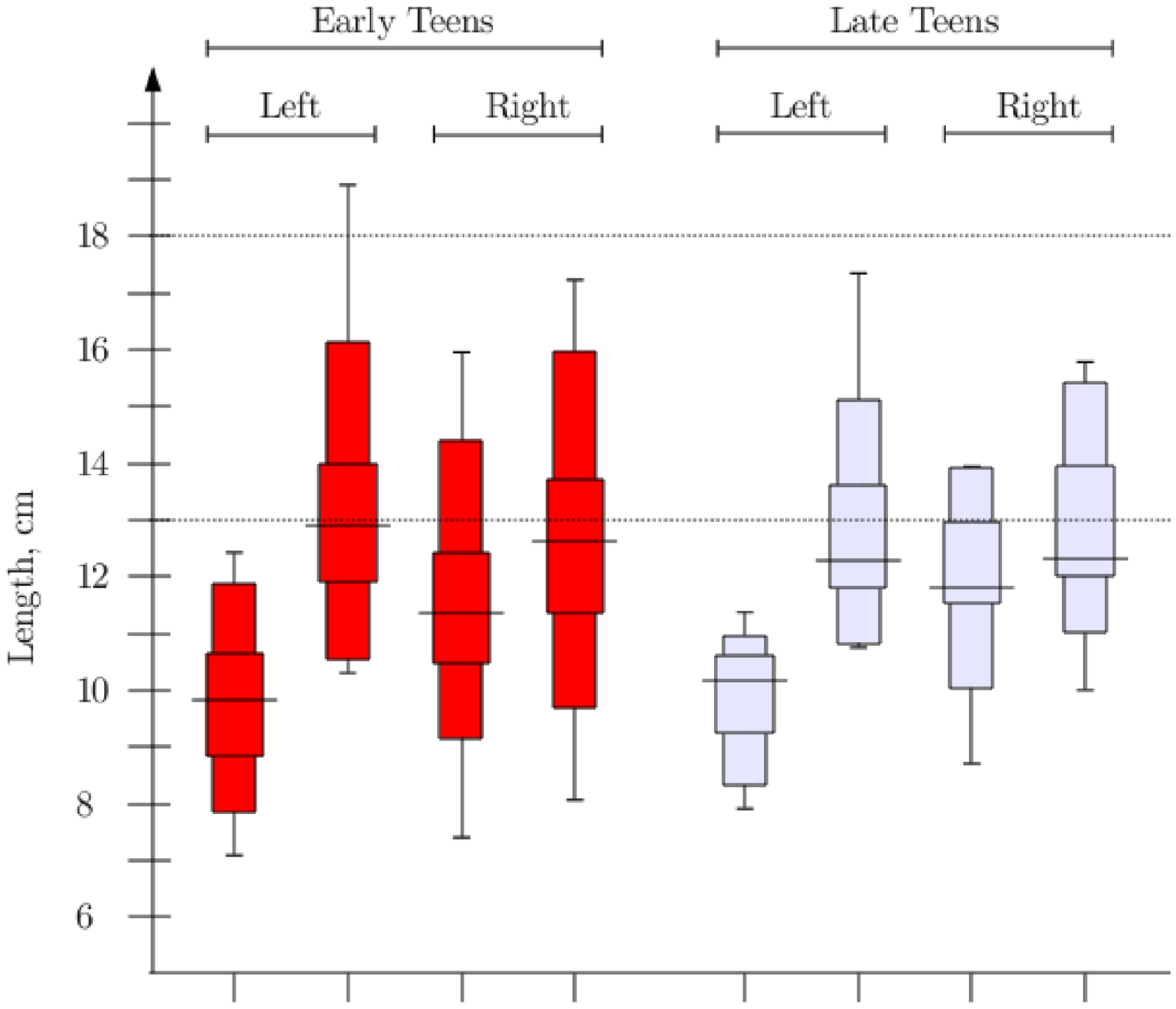, width=70mm}
\caption{Double-box and whisker plot of left and right PIR and PIE of the population by age group.}
\label{fig:data-age}
\end{figure}

Figure~\ref{fig:data-all} summarizes the measurements of left and right PIE and PRE for all samples. When the figure is compared against Table\~ref{tab:keyboard}, it is clearly seen that samples below 99th percentile will not reach the longest key combination (SHIFT--5) when the left hand is relaxed and those below 50th percentile will not be able to reach the same keys, as well as the combinations SHIFT--T and SHIFT--G when the hand is extended. Only the population above 95th percentile when the left hand is relaxed and above 25th percentile when it is extended will reach SHIFT--B (the shortest key combinations. Samples below 99th percentile will not reach the longest key combinations (SHIFT--6) when the right hand is relaxed or even extended. Samples above 95th percentile when the right hand is relaxed and above 75th percentile when the hand is extended will reach SHIFT--N and SHIFT--H. Only those above 95th percentile when the right hand is extended will reach SHIFT--Y.

%-----------------------------------------------------------------------
\subsection{Samples by Gender}
%-----------------------------------------------------------------------
Figure~\ref{fig:data-gender} shows the box and whisker plot of the samples according to the gender. The measurements shown in this figure are PIR and PIE. Comparing this figure to Table~\ref{tab:keyboard} shows that female samples below 99th percentile will not reach SHIFT--6, SHIFT--Y and SHIFT--H. Only those that fall above 99th percentile will surely reach SHIFT--N when the right hand is extended. Female samples that fall below 99th percentile will not reach SHIFT--5 when the left hand is relaxed and those that fall below 75th percentile will not reach the same keys when the hand is extended. When the left hand is extended, male samples above 25th percentile will reach SHIFT--T, SHIFT--G and SHIFT--B. Only those that fall above 95th percentile will surely reach the key combinations SHIFT--B when the same hand is relaxed. Relaxed left hand of male samples will not reach SHIFT--5, however, those above 25th percentile will reach the same keys when this hand is extended as well as keys SHIFT--T, SHIFT--G and SHIFT--B. All male samples will not reach SHIFT--6 when the right hand is either in relaxed or extended position. Samples above 95th percentile when the right hand is relaxed SHIFT--H and SHIFT--N and those above 75th percentile when the hand is extended will reach the same keys when the hand is extended.
%-----------------------------------------------------------------------
\subsection{Samples by Type of Location}
%-----------------------------------------------------------------------
Samples from urban area that fall below 99th percentile will not reach SHIFT--6 whether the hands are in relaxed or in extended position as seen in the box and whisker plot of left and right PIE and PIR of samples according to their place of origin (Figure~\ref{fig:data-location}). Similarly, those from rural area whose PIR is below 99th percentile will not reach even the shortest key combinations and the samples below 75th percentile of the extended hand will not reach the same key. Samples from urban area will that fall below 95th percentile both for extended and relaxed right hand will not reach the shortest key combinations as well. Relaxed left hand of samples both from rural and urban area will not reach SHIFT--5 while those below 75th percentile will not reach the same key combination. Only those above 99th percentile will reach SHIFT--B when the left hand is relaxed for samples from both urban and rural areas. Extended left hand of same samples that fall above 50th percentile will reach key combinations SHIFT--T, SHIFT--G and SHIFT--B

%-----------------------------------------------------------------------
\subsection{Samples by Age Group}
%-----------------------------------------------------------------------
Figure~\ref{fig:data-age} shows the box and whisker plot of the data grouped by age for PIE and PIR. Samples whose age is 17-18 are classified as early teens and those whose age ranges from 19-21 are the late teens. Early teen samples whose relaxed left hand falls below 99th percentile will not reach even the shortest key combinations (SHIFT--B), but if the same hand is extended, only the samples below 55th percentile will not reach the same keys and those below 25th percentile will not reach the other key combinations. Early teens that fall below 99th percentile will not reach SHIFT--6 when the right hand is relaxed or extended. Those below 95th percentile when the right hand is relaxed will not reach SHIT--H and SHIFT--Y and those below 95th percentile will not reach SHIFT--Y when the same hand is extended. For the late teens, those whose left PIR will not be able to reach even SHIFT-B. Those above 75th percentile will reach all key combinations the hand is extended. All late teens below 99th percentile will not reach SHIFT--6 and the relaxed right hand will not be able to reach SHIT--Y, SHIFT-H and SHIFT--N. The late teen samples whose right extended hand falls below 99th percentile will not reach SHIFT--Y and SHIFT--H and those that fall below 95th percentile will not reach SHIFT--N.

\section{Conclusion}
Hand anthropometric survey is essential especially in designing equipments and devices to minimize damage and to improve ease of use, safety and comfortability. The hand anthropometric data collected from 91 UPLB students were summarized and compared against a standard desktop keyboard and the correct placement of hands when typing. The combinations of keys with the longest range were compared against PTE. The lower percentile of the population was found out to not be able to reach the specified key combinations given the correct placement of the hand. A good keyboard design  takes into account the possible users. A 48 centimeter wide keyboard or a standard desktop keyboard is fit for most of the samples. To be able to create a keyboard design to fit most of the users, the 25th, 50th and 75th percentiles must be considered. 

Automatic hand anthropometry is a great help in the consistency of the measurements of the hand since this is done by a machine or computer. The image captured from the samples is used as an input to the program that automatically computes for the hand dimensions specified in Table~\ref{tab:handDimension}. The program outputs displayed a very minimal error when compared to the actual hand measurements. This method is more efficient when it comes to time, effort and resources like people doing the survey. Also, the machine never gets tired, unlike people do, that's why taking the measurement of the hand dimensions is more consistent. 

\section{Recommendations}
The hand anthropometric data gathered from the students can further be used towards designing mobile and other hand-held devices and other input devices. The method for automatic hand anthropometry can further be improved to further minimize the error and this can be extended to measure the lengths of other body parts and to be used for other applications.

\bibliographystyle{plainnat}
\bibliography{handAnthropometry}

\begin{thebibliography}{26}
\providecommand{\natexlab}[1]{#1}
\providecommand{\url}[1]{\texttt{#1}}
\expandafter\ifx\csname urlstyle\endcsname\relax
  \providecommand{\doi}[1]{doi: #1}\else
  \providecommand{\doi}{doi: \begingroup \urlstyle{rm}\Url}\fi

\bibitem[Amongo et~al.(2010)Amongo, Petingco, Zubia, and Jr.]{amongo10}
Rossana Marie~C. Amongo, Marvin~C. Petingco, Omar~F. Zubia, and Fernando
  O.~Paras Jr.
\newblock Empowering {F}ilipino women in agriculture through anthropometry.
\newblock In \emph{Proceedings of the 60th Philippine Society of Agricultural
  Engineers (PSAE) Annual Convention and the 21st Agricultural Engineering
  Week}. Philippine Society of Agricultural Engineers (PSAE), 2010.

\bibitem[Bailey and Pearson(1983)]{bailey83}
James~E. Bailey and Sammy~W. Pearson.
\newblock Development of a tool for measuring and analyzing computer user
  satisfaction.
\newblock \emph{Management Science}, 29\penalty0 (5):\penalty0 530--545, 1983.

\bibitem[Balakrishnan and Yeow(2007)]{vimala07}
Vimala Balakrishnan and Paul~H.P. Yeow.
\newblock {SMS} usage satisfaction influences of hand anthropometry and gender.
\newblock \emph{HUMAN IT}, 9\penalty0 (2):\penalty0 52--75, 2007.

\bibitem[Begliomini et~al.(2008)Begliomini, Nelini, Caria, Grodd, and
  Castiello]{begliomini08}
Chiara Begliomini, Cristian Nelini, Andrea Caria, Wolfgang Grodd, and Umberto
  Castiello.
\newblock Cortical activations in human grasp-related areas depend on hand used
  and handedness.
\newblock \emph{PLOS ONE}, 3\penalty0 (10):\penalty0 e3388, 2008.

\bibitem[Covavisaruch et~al.(2005)Covavisaruch, Prateepamornkul, Ruchikachorn,
  and Taksaphan]{nongluk05}
Nongluk Covavisaruch, Pipat Prateepamornkul, Puripant Ruchikachorn, and
  Piyanaat Taksaphan.
\newblock Personal verification and identification using hand geometry.
\newblock \emph{ECTI Transactions on Computer and Information Technology},
  1:\penalty0 134--140, 2005.
\newblock ECTI =.

\bibitem[Cutkosky and Howe(1990)]{cutkosky90}
Mark~R. Cutkosky and Robert~D. Howe.
\newblock Human grasp choice and robotic grasp analysis.
\newblock In Subramanian~T. Venkatamaran and Thea Iberall, editors,
  \emph{Dextrous Robot Hands}, chapter~1, pages 5--31. Springer-Verlag, 1990.

\bibitem[{DeCarlo} et~al.(1998){DeCarlo}, Metaxas, and Stone]{decarlo98}
Douglas {DeCarlo}, Dimitris Metaxas, and Matthew Stone.
\newblock An anthropometric face model using variational techniques.
\newblock In \emph{Proceedings of the 25th Annual Conference on Computer
  Graphics and Interactive Techniques (SIGGRAPH 98)}, pages 67--74, 1998.

\bibitem[{Del Prado-Lu}(2007)]{lu07}
Jinky~Leilanie {Del Prado-Lu}.
\newblock Anthropometric measurement of {F}ilipino manufacturing workers.
\newblock \emph{International Journal of Industrial Ergonomics}, 37\penalty0
  (6):\penalty0 497--503, 2007.

\bibitem[Dizmen(2012)]{dizmen12}
Coskun Dizmen.
\newblock Hand anthropometry analysis and construction of regression models for
  a {H}ong {K}ong sample.
\newblock \emph{Procedings of the International MultiConference of Engineers
  and Computer Scientists}, 2, 2012.

\bibitem[Dooley(1982)]{dooley82}
Marianne Dooley.
\newblock Anthropometric modeling programs -- {A} survey.
\newblock \emph{IEEE Computer Graphics and Applications}, 2\penalty0
  (9):\penalty0 17--25, 1982.

\bibitem[Duncan(1955)]{duncan55}
David~B. Duncan.
\newblock Multiple range and multiple {F} tests.
\newblock \emph{Biometrics}, 11\penalty0 (1):\penalty0 1--42, 1955.

\bibitem[Farkas(1994)]{farkas94}
Leslie~G. Farkas.
\newblock \emph{Anthropometry of the Head and Face}.
\newblock Raven Press, 1994.

\bibitem[Greiner(1992)]{greiner92}
T.M. Greiner.
\newblock Hand anthropometry of {U.S.} {A}rmy personnel.
\newblock Technical report, US Army Research, Development and Engineering
  Center, Natick, MA, 1992.

\bibitem[Hall et~al.(2006)Hall, Allanson, Gripp, and Slavotinek]{hall06}
Judith Hall, Judith Allanson, Karen Gripp, and Anne Slavotinek.
\newblock \emph{Handbook of Physical Measurements}.
\newblock Oxford University Press, 2nd edition, 2006.

\bibitem[Laila et~al.(2009)Laila, Ferdousi, Nurunnobi, Islam, Holy, and
  Yesmin]{zamila09}
Syeda Zamila~Hasan Laila, Roxana Ferdousi, ABM Nurunnobi, ATM~Shafiqul Islam,
  Syeda Zamila~Hosen Holy, and Farzana Yesmin.
\newblock Anthropometric measurements of the hand length and their correlation
  with the stature of {B}engali adult {M}uslim females.
\newblock \emph{Bangladesh Journal of Anatomy}, 7:\penalty0 10--13, 2009.

\bibitem[Mandahawi et~al.(2008)Mandahawi, Imrhan, Al-Shobaki, and
  Sarder]{nabeel08}
Nabeel Mandahawi, Sheik Imrhan, Salman Al-Shobaki, and B.~Sarder.
\newblock Hand anthropometry survey for the {J}ordanian population.
\newblock \emph{International Journal of Industrial Ergonomics}, 38:\penalty0
  966--976, 2008.

\bibitem[Markze(1971)]{markze71}
Mary~Walpole Markze.
\newblock Origin of the human hand.
\newblock \emph{American Journal of Physical Anthropology}, 34\penalty0
  (1):\penalty0 61--64, 1971.

\bibitem[Melcher et~al.(2003)Melcher, Sefelin, Giller, and
  Tscheligi]{melcher03}
Rudolf Melcher, Reinhard Sefelin, Verena Giller, and Manfred Tscheligi.
\newblock Improving the user experience on mobile devices and services.
\newblock In \emph{Proceedings of the Telecommunication and Mobile Computing
  Conference}, 2003.

\bibitem[{National Health and Nutrition Examination Survey}(2007)]{NHANES07}
{National Health and Nutrition Examination Survey}.
\newblock \emph{Anthropometry Procedures Manual}.
\newblock Centers for Disease Control and Prevention, 1600 Clifton Rd.,
  Atlanta, GA 30333, USA, 2007.

\bibitem[Novabos(2010)]{novabos10}
Charles~Ruel Novabos.
\newblock Anthropometric and other ergonomic interventions in the design of the
  {F}ilipino pedicab.
\newblock In \emph{First Conference of the Southeast Asian Network of
  Ergonomics Societies (SEANES)}, 2010.

\bibitem[Novabos(2012)]{novabos12}
Charles~Ruel Novabos.
\newblock The application of {F}ilipino anthropometric data on the design of
  house rooms and furniture.
\newblock In \emph{First National Conference of the Human Factors and
  Ergonomics Socity of the Philippines (HFESP)}, 2012.

\bibitem[Preedy(2012)]{preedy12}
Victor~R. Preedy, editor.
\newblock \emph{Handbook of Anthropometry: Physical Measures of Human Form in
  Health and Disease}, volume~1.
\newblock Springer, 2012.

\bibitem[Rogers(1984)]{rogers84}
Spencer~Lee Rogers.
\newblock \emph{Personal Identification from Human Remains}.
\newblock Charles C. Thomas Publisher, LTD, 1984.

\bibitem[{Timoteo-Afinidad}(2010)]{afinidad10}
Christine~B. {Timoteo-Afinidad}.
\newblock Workstation and workspace ergonomics in {P}hilippine libraries: {A}n
  emerging priority.
\newblock \emph{Journal of Philippine Librarianship}, 30\penalty0 (1):\penalty0
  21--44, 2010.

\bibitem[Ulijaszek and {Mascie-Taylor}(1994)]{ulijaszek94}
Stanley~J. Ulijaszek and C.G.~Nicholas {Mascie-Taylor}, editors.
\newblock \emph{Anthropology: The Individual and the Population (Cambridge
  Studies in Biological and Evolutionary Anthropology)}.
\newblock Cambridge University Press, 1994.

\bibitem[Wong and Pang(2005)]{wong05}
Chin~Chin Wong and Leang~Hiew Pang.
\newblock Correlations between factors affecting the diffusion of mobile
  entertainment in {M}alaysia.
\newblock In \emph{Proceedings of the 7th International Conference on
  Electronic Commerce}, pages 615--621, 2005.

\end{thebibliography}
\nocite{*}

\end{document}